# TiO$_2$ based Nanostructured Memristor for RRAM and Neuromorphic Applications: A Simulation Approach

T. D. Dongale [a, $], P. J. Patil [a], N. K. Desai [a],

P. P. Chougule [a], S. M. Kumbhar [b], P. P. Waifalkar [c], P. B. Patil [c],

R. S. Vhatkar [c], M. V. Takale [c], P. K. Gaikwad [d], R. K. Kamat [d]

[a] Computational Electronics and Nanoscience Research Laboratory,
School of Nanoscience and Biotechnology, Shivaji University, Kolhapur- 416004, India
[b] Rajarambapu Institute of Technology, Sakharale-415414, India
[c] Department of Physics, Shivaji University, Kolhapur-416004, India
[d] Embedded System and VLSI Research Laboratory, Department of Electronics,
Shivaji University, Kolhapur-416004, India

**Abstract**

We report simulation of nanostructured memristor device using piecewise linear and nonlinear window functions for RRAM and neuromorphic applications. The linear drift model of memristor has been exploited for the simulation purpose with the linear and non-linear window function as the mathematical and scripting basis. The results evidences that the piecewise linear window function can aptly simulate the memristor characteristics pertaining to RRAM application. However, the nonlinear window function could exhibit the nonlinear phenomenon in simulation only at the lower magnitude of control parameter. This has motivated us to propose a new nonlinear window function for emulating the simulation model of the memristor. Interestingly, the proposed window function is scalable up to $f(x)=1$ and exhibits the nonlinear behavior at higher magnitude of control parameter. Moreover, the simulation results of proposed nonlinear window function are encouraging and reveals the smooth nonlinear change from LRS to HRS and vice versa and therefore useful for the neuromorphic applications.

[$] **Corresponding Author:** T. D. Dongale

E-mail: tukaram.eln@gmail.com

# 1. Introduction

Memristor which is poised to establish as the fourth circuit element in addition to the R, L and C, was theorized way back in the year 1971 by Prof. L. Chua [1]. Later in the year 2008 the same was validated by the HP research group [2]. The peculiar characteristics of remembering the data in terms of Low Resistance State (LRS) and High Resistance State (HRS) makes the memristor a unique attribute for many interesting applications not feasible with the conventional circuit elements. Moreover, the passivity and nonlinearity are some of the important characteristics of the memristor, which leads to its usage in the applications of diversified domains such as biomedical, Resistive Random Access Memory (RRAM), neural computing, nonlinear dynamics, neuromorphic computing realm etc as reported widely in the literature [3-10]. As these applications are important so is the accurate modelling of nonlinear memristor which has been the very basis of the scientific investigations. Incidentally many research group including ours are actively working in this direction as put forth briefly in the following paragraph to set the background of the present investigation.

Recently, Li et al reported a new modelling method with multinomial window function. This method is derived through the statistical fitting of an experimental data of a memristor device [11]. Batas et al have come out with the behavioral model of Magnetic Flux-Controlled memristor device. The reported model is simulated on integrated circuits emphasis i.e. SPICE platform [12]. Valsa et al have investigated the analogue model of the memristor device which has been duly verified for various test signals with the results showing good resemblance with the ones reported in literature [13]. Shin et al have put forth a compact circuit model and hardware emulator for memristor device with its applications for the arithmetic operations [14]. Kolka et al reported the hardware emulator for the mem-systems based on the memristor, memcapacitor, and meminductor which can be further programmed to realize the above mentioned trio [15]. Quite relevant to the theme of present paper are the investigations carried out by Biolek et al and Yu et al on the nonlinear and piecewise linear window function aspects for modelling the memristor respectively [16-17].

Recently, our research group too has reported two new window functions for the modeling of the memristor device [18]. The present research paper is an extension of our previously reported work [3-10, 18]. While our previous papers report more of the details related to mathematical aspects, the present paper is a value addition as it actually showcases the simulation in light of the RRAM and neuromorphic applications. Both these applications are currently in profound demand. RRAM's seems to be the only solution in the age of big data while a completely new paradigm of brain inspired computing is currently been explored through the neuromorphic domain. The main achievement of the present manuscript is the modified nonlinear window function which accurately models the nonlinearity of memristor device. The rest of the paper is as follows, after brief introduction in the first section, second section deals with the overview of piecewise linear and nonlinear window functions. The third section further divulges the simulation details of memristor with above mentioned window function. This section also deals with the modified nonlinear window function. At the end results and conclusion has been placed.

## 2. Overview of piecewise linear and nonlinear window functions

The HP research group modeled the memristor device based on linear drift model. This model assumes that the state variable (w) is directly proportional to charges flowing through the device [2]. The structure of HP memristor (Pt/TiO$_2$/Pt memristor) is shown in fig. 1. It has two prominent regions namely doped low resistance and undoped high resistance.

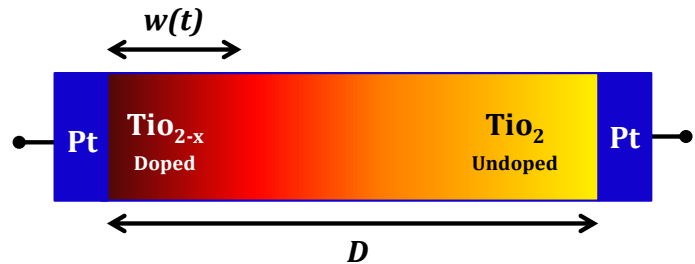

**Fig. 1**. Structure of memristor reported by HP research group [2, 18]

The reported literature reveals that drifting of vacancies has been highly nonlinear near the boundary interfaces. This is attributed to the nanoscale phenomena by which even a small voltage can produce large electric field across the device. This large electric field further generates nonlinear drifting of vacancies near the boundary interfaces [19]. Another problem with linear drift model of memristor is that, the state variable 'w' never reaches to zero physical length which indicates that the oxygen vacancies are absent in the devices [20]. The boundary problem can be minimized by adopting window function *f(x)*. In general, the window function can be multiplied to state equation of memristor which is given as,

$$\frac{dw(t)}{dt} = \eta \frac{\mu_v R_{ON}}{D} i(t) * f(x) \qquad (1)$$

where, *w* is a state variable, the parameter $\eta$ indicates the polarity of memristor e.g. $\eta=1$ indicates the expansion of doped region and $\eta=-1$ indicates the shrinking of doped region, $\mu_V$ is a average drift velocity of oxygen vacancies, $R_{ON}$ is a low resistance state of memristor device or ON state resistance, D corresponds to total length of the active region, and *i* is a current through the device. The function *f(x)* should have its highest value at the center of the device (x=0.5) and zero at the boundaries (x=0 and x=1) of memristor device [20]. From the physics point of view, window functions lower down the speed of oxygen vacancies near the boundaries which ultimately leads to the nonlinear behavior.

In the backdrop of the above theoretical propositions, it is appropriate to select the fitting window functions most suited to the intended applications to the core of our research group. With the RRAM and neuromorphic domain of applications, piecewise linear and nonlinear window functions have been applied owing to their benefits in terms of parameter adjustment flexibility. Upon applying, the piecewise linear window function exhibits continuously differentiability at LHS bounds, middle region and RHS bounds. It shows the nonlinear behavior at lower values of control parameter '*p*' and linear behavior at higher values of control parameter '*p*'. The nonlinear window function shows the quasi nonlinear behavior at higher values of control parameter '*p*'. One of the advantages of these window functions is that the control parameter '*p*' can be adjusted to get required characteristics of memristor pertaining to RRAM and neuromorphic applications [18]. Equation 2 represents the generalized piecewise linear window function such that, [18]

$$f(x) = \begin{cases} \dfrac{px}{mX_0} & \text{for } 0 \leq x \leq X_0 \\ \dfrac{p}{m} & \text{for } X_0 \leq x \leq Y_0 \\ \dfrac{p(1-x)}{m(1-Y_0)} & \text{for } Y_0 \leq x \leq 1 \end{cases} \qquad (2)$$

where, $0 < X_0 < Y_0 < 1$ and p and m ∈ R⁺. The nonlinear window function has similar characteristics with respect to piecewise linear window function except, it has nonlinear characteristics at the boundaries. Equation 3 represents the nonlinear window function such that, [18]

$$f(x) = \begin{cases} x^{\frac{1}{p}} & \text{for } 0 \leq x \leq X_0 \\ x_0^{\frac{1}{p}} & \text{for } X_0 \leq x \leq Y_0 \\ \left|(x-1)^{\frac{1}{p}}\right| & \text{for } Y_0 \leq x \leq 1 \end{cases} \qquad (3)$$

where, $0 < X_0 < Y_0 < 1$, $Y_0 = (1 - X_0)$ and p ∈ R⁺. Fig. 2 represents the piecewise linear and nonlinear window function with various values of control parameter *'p'*.

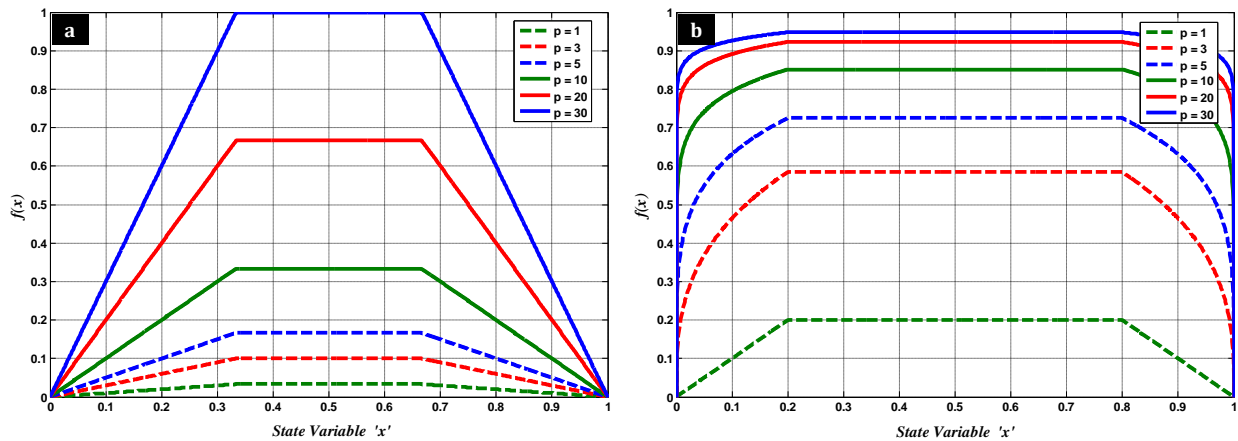

**Fig. 2**. Piecewise linear and nonlinear window functions with various values of control parameter *'p'* and state variable location.

## 3. Simulation of memristor device using piecewise linear and nonlinear window functions

The main intent behind the modeling and simulation is to help the designers to come out with the apt device characteristics per application. In the present case the main rationale is to fine tune the memristor attributes through simulation for two fold purposes viz. fast transition from LRS to HRS for RRAM applications while slow transition from LRS to HRS for the neuromorphic domain. The modeling for the above mentioned attributes has been obtained by applying the piecewise linear and nonlinear window functions. After zeroing down on the technique for modeling the simulation was accomplished.

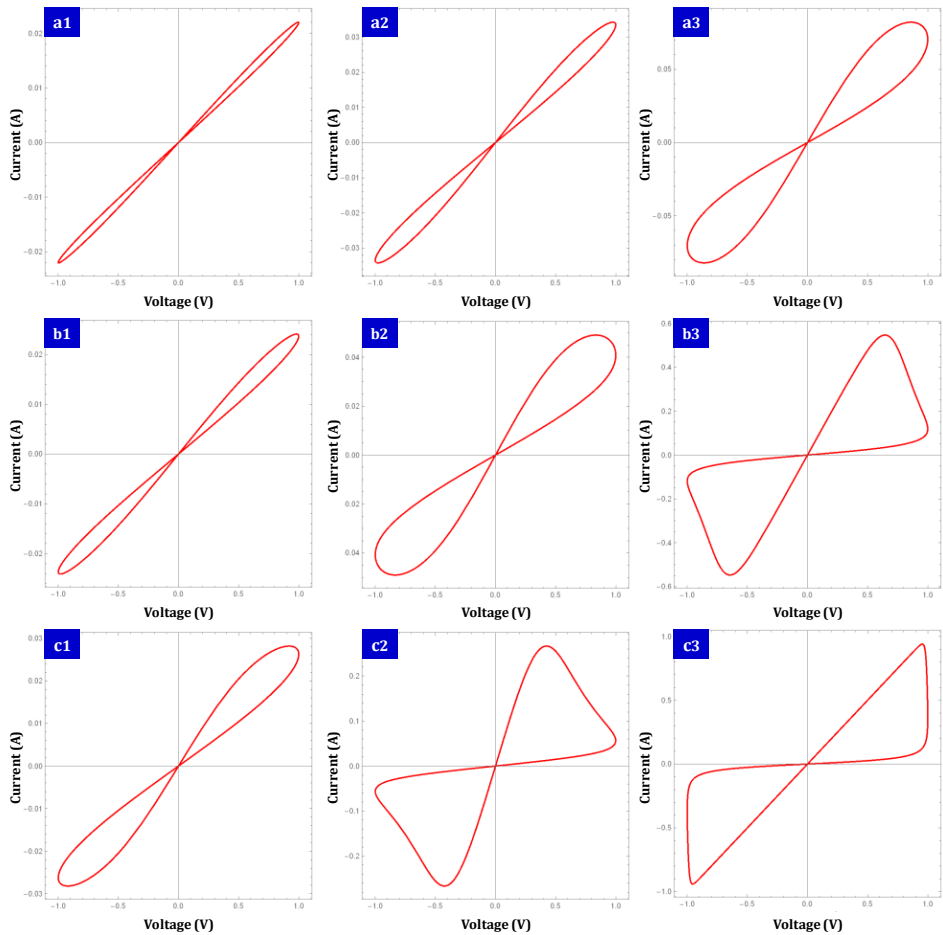

**Fig. 3:** I-V characteristics of nanostructured memristor device with piecewise linear window function. Fig. (a1 to a3) represents I-V characteristics of memristor device at x=0.3, x=0.5, and x=0.7 respectively with control parameter p=10. Fig. (b1 to b3) ) represents I-V characteristics of memristor device at x=0.3, x=0.5, and x=0.7 respectively with control parameter p=20. Fig. (c1 to c3) ) represents I-V characteristics of memristor device at x=0.3, x=0.5, and x=0.7 respectively with control parameter p=30.

Fig. 3 represents the simulated I-V characteristics of nanostructured memristor device with piecewise linear window function. The present simulation is carried out for control parameter p = 10, 20, and 30 with state variable x = 0.3, 0.5, and 0.7. In the other words, these state variables represent the growth location of conductive filament in the memristor device. An illustration of such mechanism is shown in the fig. 4. The state variable *x*=0.3, 0.5, and 0.7 represents LHS bounds, middle region and RHS bounds respectively and it can be visualize from fig. 2. The simulation results suggested that the current in the device increases as the control parameter *p* increases from 10 to 30. The relationship between current and control parameter with various values of state variable is shown in the fig. 5. From fig. 3 and 5, it is seen that the memristor device shows RRAM kind of characteristics at higher magnitude of control parameter. The results also suggest that abrupt switching occurs at the higher magnitude of control parameter. For the lower magnitude of control parameter, current and pinched hysteresis loop (PHL) become small. The area under the PHL is also increases as the magnitude of control parameter increases. The change in the control parameter can be used for the switching from one state to another state. In the other words, if one can have power over the control parameter then switching of the device can be controlled. This characteristic is very similar to digital memory and has application in the digital memory domain. From the results it is clearly evident that memristor will be work as a promising RRAM building block at the higher values of control parameter, when it is modeled with piecewise linear window function.

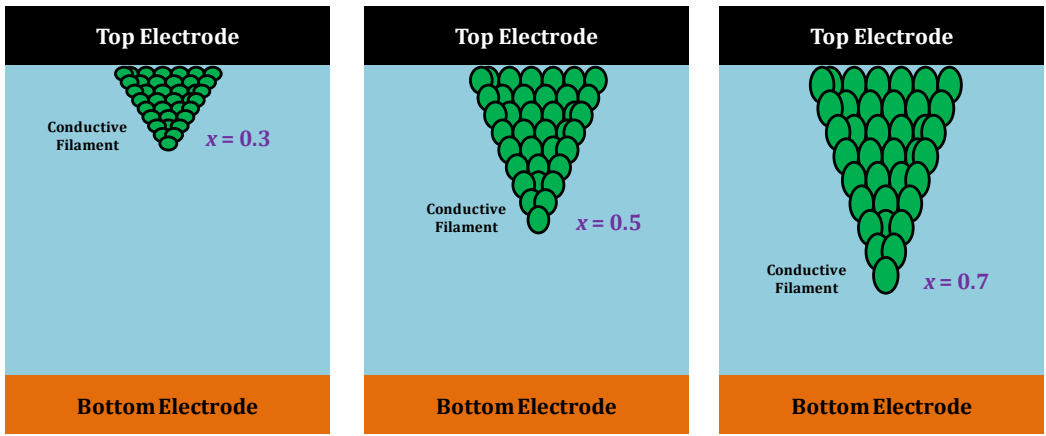

**Fig. 4**. Illustration of growth location of conductive filament in the memristor device. The state variable x = 0.3, 0.5, and 0.7 represents LHS bounds, middle region and RHS bounds respectively.

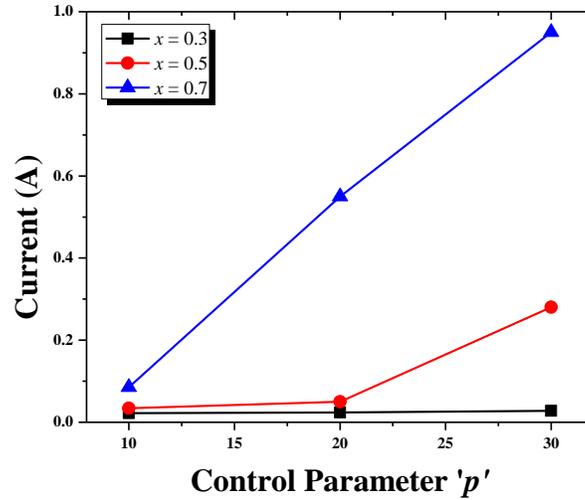

**Fig. 5.** The relationship between current and control parameter with various values of state variable (*x*). It is clear that current abruptly increases at higher value of control parameter.

Fig. 6 represents the simulated I-V characteristics of nanostructured memristor device with nonlinear window function. For the present simulation control parameter 'p' varies as 1, 3, 5, 10, 20, and 30. For each control parameter, I-V characteristics is simulated at state variable *x*=0.5 (insignificant change is observed at state variable *x*=0.3 and 0.7 with respect to *x*=0.5).

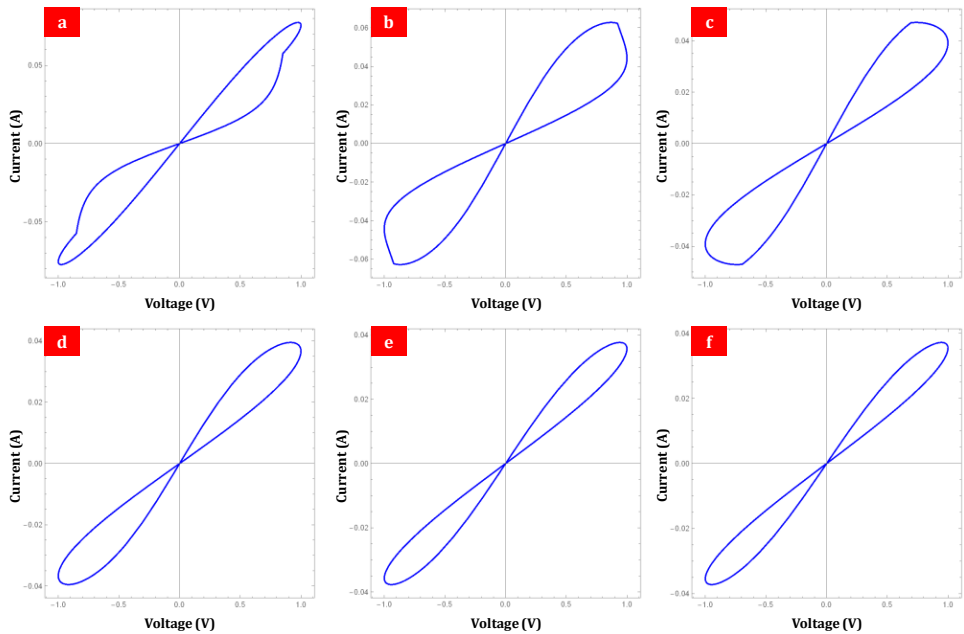

**Fig. 6:** I-V characteristics of nanostructured memristor device with nonlinear window function. Fig. (a to f) represents the I-V characteristics of memristor device at x=0.5 and control parameter p=1, 3, 5, 10, 20, and 30 respectively.

The results indicates that the memristor device shows nonlinear behavior only at the lower magnitude of control parameter (p=1, 3, and 5) and I-V characteristics does not alter at higher magnitude of control parameter (p=10, 20, and 30). The area under the curve is higher only at the lower magnitude of control parameter and becomes approximately same at higher magnitude of control parameter. This window function does not reaches to $f(x)=1$ (it becomes $f(x)=1$ only at p=∞) and is a main limitation of nonlinear window function. To rectify this limitation, we are proposing a new window function which can be scaled up to $f(x)=1$ at higher magnitude of control parameter. The proposed nonlinear window function can be defined as,

If $p < p_0$ then,

$$f(x) = \begin{cases} x^{\frac{1}{p}} & \text{for } 0 \leq x \leq X_0 \\ x_0^{\frac{1}{p}} & \text{for } X_0 \leq x \leq Y_0 \\ \left|(x-1)^{\frac{1}{p}}\right| & \text{for } Y_0 \leq x \leq 1 \end{cases}$$

Otherwise if $p = p_0$ then, (4)

$$f(x) = \begin{cases} x^{\left(\frac{1-x_0 x}{p}\right)} & \text{for } 0 \leq x \leq X_0 \\ x_0^{\frac{p_0-p}{p}} & \text{for } X_0 \leq x \leq Y_0 \\ |(x-1)|^{-\left(\frac{x_0 x - 1}{p}\right)} & \text{for } Y_0 \leq x \leq 1 \end{cases}$$

where, $0 < X_0 < Y_0 < 1$, $Y_0 = (1- X_0)$ and $p \in R^+$. Fig. 7 shows the difference between two nonlinear window functions with various values of control parameter 'p'. The results suggested that the proposed window function scaled up to $f(x)=1$ at higher magnitude of control parameter. The simulation of memristor device with modified nonlinear window function is shown in the fig. 8. The results suggested that modified nonlinear window function is able to simulate the memristor characteristics at higher magnitude of control parameter. From the results it is clear that the current in the device increases as a function of state variable i.e. magnitude of current increases as value of state variable increases. I-V characteristic shows smooth nonlinear change from LRS to HRS and vice versa. This is very similar to analog memory and has application in the neuromorphic engineering domain. In nutshell, modified nonlinear window function can be used for the analog memory applications.

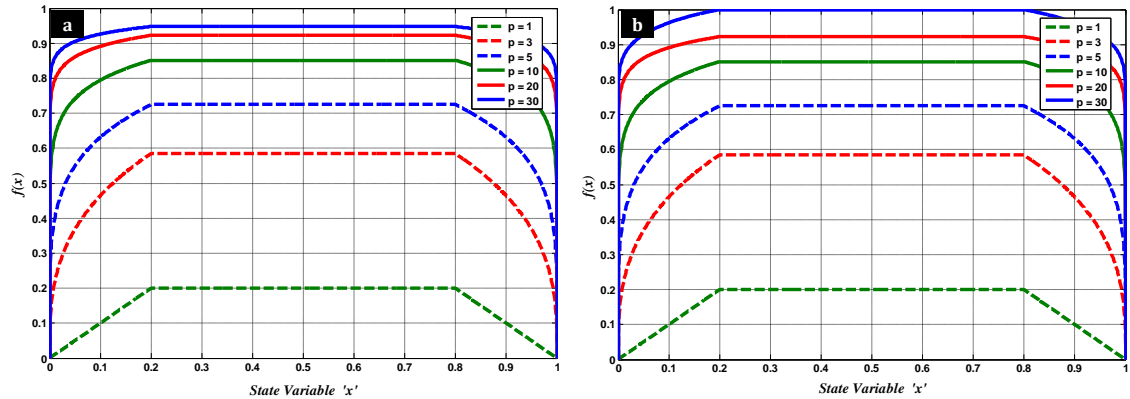

**Fig. 7.** Difference between two nonlinear window functions with various values of control parameter *'p'*.

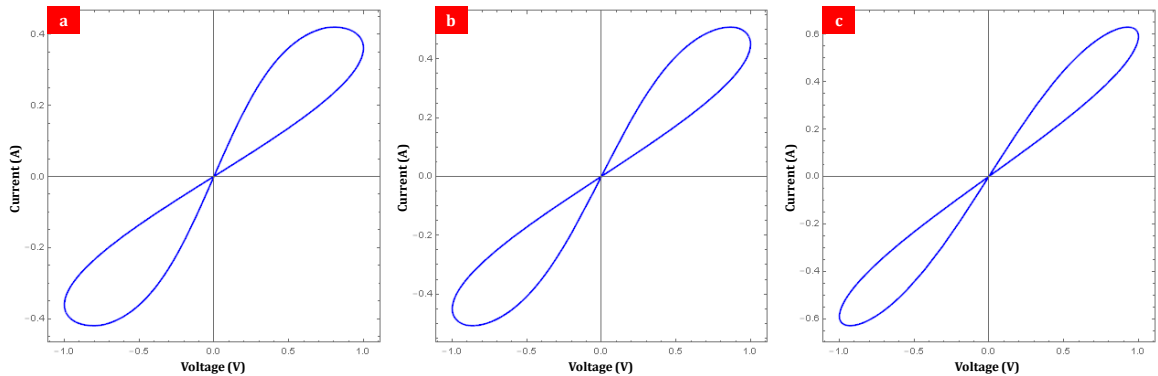

**Fig. 8.** (a-c) Simulation of memristor device with modified nonlinear window function at control parameter p=30 and state variable *x*=0.3, 0.5 and 0.7 respectively.

## 4. Conclusion

The present investigation reports the simulation of TiO$_2$ nanostructured memristor device for RRAM and neuromorphic applications. The results strongly indicate the suitability of piecewise linear window function to carve the model of the nanostructured memristor device characteristics for RRAM application which is further been validated through simulation. Altering the control parameter from one state to another state makes the piecewise linear window function a best fit for the RRAM application. The modified nonlinear window function eliminates the scaling issue and thus accomplishes simulation of the memristor characteristics at higher magnitude of control parameters. The results are encouraging and show strong applicability towards neuromorphic engineering domains on which our research investigations are in progress.

**References**

1) L. O. Chua, *IEEE Transactions on Circuit Theory, 18*(5), (1971), 507-519.
2) D. B. Strukov, G. S. Snider, D. R. Stewart, R. S. Williams, *Nature*, 453(7191), (2008), 80-83.
3) T. D. Dongale, *Int. J. Health Informatics*, 2(1), (2013), 15-20.
4) T. D. Dongale, S. S. Shinde, R. K. Kamat, K. Y. Rajpure, *J. Alloy and Compounds*, 593, (2014), 267-270.
5) T. D. Dongale, K. P. Patil, S. R. Vanjare, A. R. Chavan, P. K. Gaikwad, R. K. Kamat, *Journal of Computational Science*, 11, (2015), 82–90.
6) T. D. Dongale, S. V. Mohite, A. A. Bagade, P. K. Gaikwad, P. S. Patil, R. K. Kamat, K. Y. Rajpure, *Electronic Materials Letters*, 11(6), (2015), 944-948.
7) T. D. Dongale, K. P. Patil, P. K. Gaikwad, R. K. Kamat, *Materials Science in Semiconductor Processing*, 38, (2015), 228-233.
8) T. D. Dongale, K. V. Khot, S. S. Mali, P. S. Patil, P. K. Gaikwad, R. K. Kamat, P. N. Bhosale, *Material Science in Semiconductor Processing*, 40, (2015), 523–526.
9) T. D. Dongale, K. P. Patil, S. B. Mullani, K. V. More, S. D. Delekar, P. S. Patil, P. K. Gaikwad, R. K. Kamat, *Materials Science in Semiconductor Processing,* 35, (2015), 174-180.
10) S. S. Shinde, and T. D. Dongle, *Journal of Semiconductors*, 36(3), (2015), 034001-3.
11) G. Li, J. Mathew, R. Shafik & D. Pradhan, *International Journal of Electronics Letters*, 3(1), (2015), 1-12.
12) D. Batas, & H. Fiedler, *IEEE Transactions on Nanotechnology*, 10(2), (2011), 250-255.
13) J. Valsa, D. Biolek, & Z. Biolek, *International Journal of Numerical Modelling: Electronic Networks, Devices and Fields*, 24(4), (2011), 400-408.
14) S. Shin, L. Zheng, G. Weickhardt, S. Cho & S. M. Kang, IEEE Circuits and Systems Magazine, 13(2), (2013), 42-55.
15) Z. Kolka, D. Biolek, & V. Biolková, *International Journal of Numerical Modelling: Electronic Networks, Devices and Fields*, 25(3), (2012), 216-225.
16) Z. Biolek, D. Biolek, & V. Biolkova, *Radioengineering*, 18(2), (2009), 210-214.
17) J. Yu, X. Mu, X. Xi & S. Wang, *Radioengineering*, 22(4), (2013), 969-974.
18) T. D. Dongale, P. J. Patil, K. P. Patil, S. B. Mullani, K. V. More, S. D. Delekar, P. K. Gaikwad, R. K. Kamat, *Journal of Nano- and Electronic Physics*, 7(3), (2015), 03012-1-03012-4.
19) J. J. Yang, M. D. Pickett, X. Li, D. A. A. Ohlberg, D. R. Stewart, and R. S. Williams, *Nature Nanotechnology*, 3(7), (2008), 429–433.
20) N. R. McDonald, R. E. Pino, P. J. Rozwood and B. T. Wysocki. IEEE International Joint Conference on Neural Networks (IJCNN), 1-5, 2010.